\documentclass{article}
\usepackage{spconf,amsmath,epsfig}
\usepackage{amsfonts}
\usepackage{url}

\title{Towards Deep Unsupervised SAR Despeckling with Blind-Spot Convolutional Neural Networks}

\name{Andrea Bordone Molini, Diego Valsesia, Giulia Fracastoro, Enrico Magli\thanks{This research has been funded by the Smart-Data@PoliTO center
for Big Data and Machine Learning technologies.}}
\address{Politecnico di Torino, Italy}
\begin{document}
%
\maketitle
\begin{abstract}
SAR despeckling is a problem of paramount importance in remote sensing, since it represents the first step of many scene analysis algorithms. Recently, deep learning techniques have outperformed classical model-based despeckling algorithms.
However, such methods require clean ground truth images for training, thus resorting to synthetically speckled optical images since clean SAR images cannot be acquired. In this paper, inspired by recent works on blind-spot denoising networks, we propose a self-supervised Bayesian despeckling method. The proposed method is trained employing only noisy images and can therefore learn features of real SAR images rather than synthetic data. We show that the performance of the proposed network is very close to the supervised training approach on synthetic data and competitive on real data.
\end{abstract}
\vspace*{-6pt}
\section{Introduction}
\label{sec:intro}
\vspace*{-6pt}
Synthetic Aperture Radar (SAR) is a coherent imaging system and as such it strongly suffers from the presence of speckle, a signal dependent granular noise. Speckle noise makes SAR images difficult to interpret, preventing the effectiveness of scene analysis algorithms for, e.g., image segmentation, detection and recognition.
Several despeckling methods applied to SAR images have been proposed working either in spatial or transform domain. 
The first attempts at despeckling employed filtering-based techniques operating in spatial domain such as Lee filter \cite{LEE198124}, Frost filter \cite{4767223}, Kuan filter \cite{1165131}, and Gamma-MAP filter \cite{doi:10.1080/01431169308953999}.
Wavelet-based methods \cite{1105905,1166595} enabled multi-resolution analysis. More recently, non-local filtering methods attempted to exploit self-similarities and contextual information. A combination of non-local approach, wavelet domain shrinkage and Wiener filtering in a two-step process led to SAR-BM3D \cite{5989862}, a SAR-oriented version of BM3D \cite{4271520}.

In recent years, deep learning techniques have set the benchmark in many image processing tasks, achieving exceptional results in problems such as image restoration \cite{7839189}, super resolution \cite{deepsum}, semantic segmentation \cite{7298965}.
Recently, some despeckling methods based on convolutional neural networks (CNNs) have been proposed \cite{8053792,CozzolinoCNN}, attempting to leverage the feature learning capabilities of CNNs. Such methods use a supervised training approach where the network weights are optimized by minimizing a distance metric between noisy inputs and clean targets. 
However, clean SAR images do not exist and supervised training methods resort to synthetic datasets where optical images are used as ground truth and their artificially speckled version as noisy inputs. This creates a domain gap between the features of synthetic training data and those of real SAR images, possibly leading to presence of artifacts or poor preservation of radiometric features. SAR-CNN \cite{CozzolinoCNN} addressed this problem by averaging multi-temporal SAR data of the same scene to obtain a ground truth. However, acquisition of multi-temporal data, scene registration and robustness to variations can be challenging.

Self-supervised denoising methods represent an alternative to train CNNs without having access to the clean images.
Noise2Noise \cite{pmlr-v80-lehtinen18a} proposed to use pairs of images with the same content but independent noise realizations. This method is not suitable for SAR despeckling due to the difficulty in accessing multiple images of the same scene with independently drawn noise realizations.
Noise2void \cite{Krull2018Noise2VoidL} further relaxes the constraints on the dataset, requiring only a single noisy version of the training images, by introducing the concept of blind-spot networks. Assuming spatially uncorrelated noise, and excluding the center pixel from receptive field of the network, the network learns to predict the value of the center pixel from its receptive field by minimizing the $\ell_2$ distance between the prediction and the noisy value. The network is prevented from learning the identity mapping because the pixel to be predicted is removed from the receptive field. The blind-spot scheme used in Noise2void \cite{Krull2018Noise2VoidL} is carried out by a simple masking method, keeping a few pixels active in the learning process.
Laine et al. \cite{laine2019high} devised a novel convolutional blind-spot network architecture capable of processing the entire image at once, increasing the efficiency.
They also introduce a Bayesian framework to include noise models and priors on the conditional distribution of the blind spot given the receptive field.

In this paper, we use the self-supervised Bayesian denoising with blind-spot networks proposed in \cite{laine2019high}, adapting the model to the noise and image statistics of SAR images, thus enabling direct training on real SAR images. 
Our method bypasses the problem of training a CNN on synthetically-speckled optical images and using it to denoise SAR images, since in general transfer knowledge from optical to SAR images is a very difficult task as imaging geometries and content are quite dissimilar due to the different imaging mechanisms.
To the best of our knowledge, this is the first self-supervised method to deal with real SAR images.

\vspace*{-8pt}
\section{Background}
\label{sec:background}
\vspace*{-8pt}

CNN denoising methods estimate the clean image by learning a function that takes each noisy pixel and combines its value with the local neighboring pixel values (receptive field) by means of multiple convolutional layers interleaved with non-linearities. Taking this from a statistical inference perspective, a CNN is a point estimator of $p(x_i|y_i,\Omega_{y_i})$, where $x_i$ is the $i^{th}$ clean pixel, $y_i$ is the $i^{th}$ noisy pixel and $\Omega_{y_i}$ represents the receptive field composed of the noisy neighboring pixels, excluding $y_i$ itself.
Noise2void predicts the clean pixel $x_i$ by relying solely on the neighboring pixels and using $y_i$ as a noisy target. The CNN learns to produce an estimate of $\mathbb{E}_{x_i}[x_i|\Omega_{y_i}]$ using the $\ell_2$ loss when in presence of Gaussian noise.
The drawback of Noise2void is that the value of the noisy pixel $y_i$ is never used to compute the clean estimate. 

The Bayesian framework devised by Laine et al. \cite{laine2019high} explicitly introduces the noise model $p(y_i|x_i)$ and conditional pixel prior given the receptive field $p(x_i|\Omega_{y_i})$ as follows: 
\vspace{-8pt}
\begin{align*}
p(x_i|y_i,\Omega_{y_i}) \propto p(y_i|x_i) p(x_i|\Omega_{y_i}).\\[-20pt]
\end{align*}
The role of the CNN is to predict the parameters of the chosen prior $p(x_i|\Omega_{y_i})$. The denoised pixel is then obtained as the MMSE estimate, i.e., it seeks to find $\mathbb{E}_{x_i}[x_i|y_i,\Omega_{y_i}]$.
Under the assumption that the noise is pixel-wise i.i.d., the CNN is trained so that the data likelihood $p(y_i|\Omega_{y_i})$ for each pixel is maximized. The main difficulty involved with this technique is the definition of a suitable prior distribution that, when combined with the noise model, allows for close-form posterior and likelihood distributions. We also remark that while imposing a handcrafted distribution as $p(x_i|\Omega_{y_i})$ may seem very limiting, it is actually not since i) that is the \textit{conditional} distribution given the receptive field rather than the raw pixel distribution, and ii) its hyperparameters are predicted by a powerful CNN on a pixel-by-pixel basis. 

\vspace*{-8pt}
\section{Proposed method}
\label{sec:method}
\vspace*{-8pt}

\begin{figure}
\centering
\includegraphics[width=0.9\linewidth]{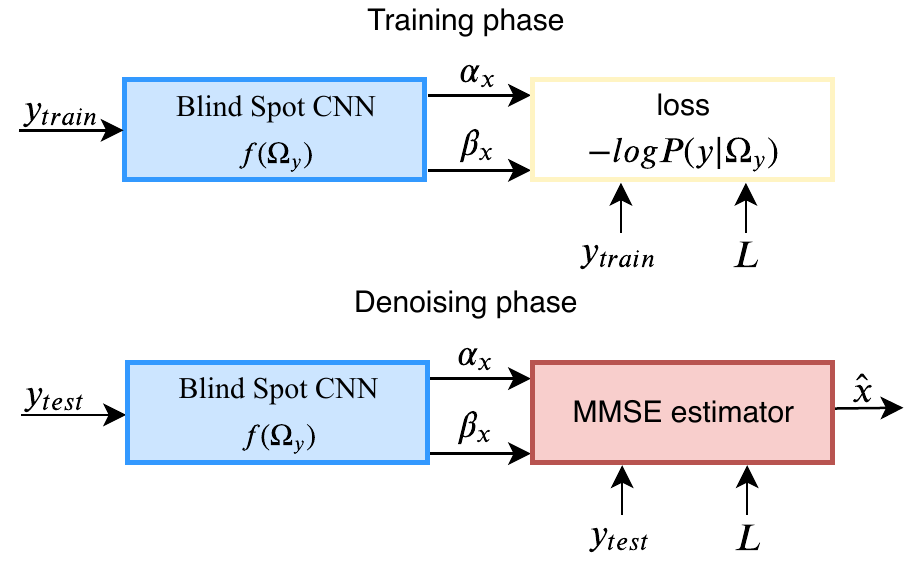}
\vspace{-0.4cm}
\caption{Scheme depicting the training and the testing phases.}
\vspace{-0.3cm}
\label{fig:summary}
\end{figure}

Following the notation in Sec. \ref{sec:background}, this section presents the Bayesian model we adopt for SAR despeckling and the training procedure. A summary is shown in Fig. \ref{fig:summary}. 

\vspace*{-8pt}
\subsection{Model}
\vspace*{-5pt}
We consider the multiplicative SAR speckle noise model: $y_i = n_i x_i$ where $x$ represents the unobserved clean image and $n$ the uncorrelated multiplicative speckle. Concerning noise modeling, we choose the widely-used $\Gamma(L,L)$ distribution for an $L$-look image.
We model the conditional prior distribution given the receptive field as an inverse Gamma distribution with shape $\alpha_{x_i}$ and scale $\beta_{x_i}$:
\begin{align*}
    p(x_i|\Omega_{y_i}) = \mathrm{inv}\Gamma(\alpha_{x_i},\beta_{x_i}),
\end{align*}
where $\alpha_{x_i}$ and $\beta_{x_i}$ depend on $\Omega_{y_i}$, since they are the outputs of the CNN at pixel $i$.
For the chosen prior and noise models, the posterior distribution is also an inverse Gamma:
\begin{align}\label{eq:posterior}
    p(x_i|y_i,\Omega_{y_i}) = \mathrm{inv}\Gamma(L+\alpha_{x_i},\beta_{x_i}+Ly_i).
\end{align}

Finally, the noisy data likelihood $p(y_i|\Omega_{y_i})$ can be obtained in closed form:
\begin{align*}
p(y_i|\Omega_{y_i})&=\frac{L^L y^{L-1}_i}{\beta^{-\alpha_{x_i}}_{x_i} Beta(L,\alpha_{x_i}) (\beta_{x_i} + L y_i)^{L+\alpha_{x_i}}},
\end{align*}
with the Beta function defined as $Beta(L, \alpha_{x_i}) = \frac{\Gamma(L)\Gamma(\alpha_{x_i})}{\Gamma(L+\alpha_{x_i})}$.
This distribution is also known as the $G^0_I$ distribution introduced in \cite{581981}. It has been observed that it is a good model of highly heterogeneous SAR data in intensity format like urban areas, primary forests and a deforested area.

\vspace*{-8pt}
\subsection{Training}
\vspace*{-5pt}

The training procedure learns the weights of the blind-spot CNN, which is used to produce the estimates for parameters $\alpha_{x_i}$ and $\beta_{x_i}$ of the inverse gamma distribution $p(x_i|\Omega_{y_i})$. We refer the reader to \cite{laine2019high} on how to implement a CNN so that it has a central blind spot.
The blind-spot CNN is trained to minimize the negative log likelihood $p(y_i|\Omega_{y_i})$ for each pixel, so that the estimates of $\alpha_{x_i}$ and $\beta_{x_i}$ fit the noisy observations. Our loss function is as follows:
\begin{align*}
l &= - \sum_{i} \log p(y_{i}|\Omega_{y_{i}}).
\end{align*}

\vspace*{-10pt}
\subsection{Testing}
\vspace*{-6pt}

In testing, the blind-spot CNN processes the SAR image to estimate $\alpha_{x_i}$ and $\beta_{x_i}$ for each pixel. 
The despeckled image is then obtained through the MMSE estimator, i.e., the expected value of the posterior distribution in Eq. \eqref{eq:posterior}:\vspace*{-4pt}
\begin{align*}
    \hat{x}_i = \mathbb{E}[x_i|y_i,\Omega_{y_i}] = \frac{\beta_{x_i} + L y_i}{L+\alpha_{x_i}-1}.\\[-16pt]
\end{align*}
Notice that this estimator combines both the per-pixel prior estimated by the CNN and the noisy realization.

\begin{table}
\centering
\caption{Synthetic images - PSNR (dB)}
\label{table:synth_images}
\resizebox{\columnwidth}{!}{%
\begin{tabular}{lcccc}
Image & PPB \cite{5196737} & SAR-BM3D \cite{5989862} & SAR-CNN \cite{CozzolinoCNN} & Proposed \\ \hline
Cameraman & 23.02 & 24.76 & 26.15 & 25.90 \\ \hline
House & 25.51 & 27.55 & 28.60 & 27.96 \\ \hline
Peppers & 23.85 & 24.92 & 26.02 & 25.99 \\ \hline
Starfish & 21.13 & 22.71 & 23.37 & 23.32\\ \hline
Butterfly & 22.76 & 24.48 & 26.05 & 25.82 \\ \hline
Airplane & 21.22 & 22.71 & 23.93 & 23.67 \\ \hline
Parrot & 21.88 & 24.17 & 25.92 & 25.44 \\ \hline
Lena & 26.64 & 27.85 & 28.70 & 28.54 \\ \hline
Barbara & 24.08 & 25.37 & 24.70 & 24.36 \\ \hline
Boat & 24.22 & 25.43 & 26.05 & 26.02 \\ \hline
\textit{Average} & \textit{23.43} & \textit{24.99} & \textit{25.95} & \textit{25.67}\\ \hline

\end{tabular}%
}
\vspace*{-12pt}
\end{table}

\begin{table}
\centering
\caption{Quantitative results on SAR real images}
\label{table:real_images}
\resizebox{\columnwidth}{!}{%
\begin{tabular}{lcccc}
Metrics & PPB \cite{5196737} & SAR-BM3D \cite{5989862} & SAR-CNN \cite{CozzolinoCNN} & Proposed \\ \hline
 $\mu_r$ & 1.0021 & 1.0628 &  0.9845 & 1.0271 \\ \hline
 $\sigma_r$ & 1.4004 & 1.7322 & 0.8458 & 0.9837 \\ \hline
 ENL & 44.56 & 22.80 & 29.98 & 8.91 \\ \hline
\end{tabular}%
}
\vspace*{-12pt}
\end{table}


\begin{figure*}[t]
  \centering
    \begin{minipage}[b]{\textwidth}
        \begin{minipage}[c]{0.18\textwidth}
        \includegraphics[width=\textwidth]{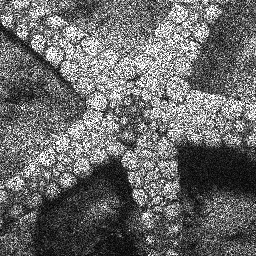}
        \end{minipage}
        \hfill
        \begin{minipage}[c]{0.18\textwidth}
        \includegraphics[width=\textwidth]{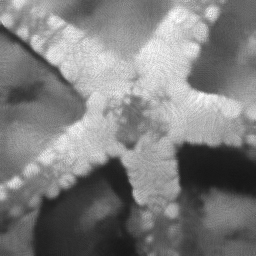}
        \end{minipage}
        \hfill
        \begin{minipage}[c]{0.18\textwidth}
        \includegraphics[width=\textwidth]{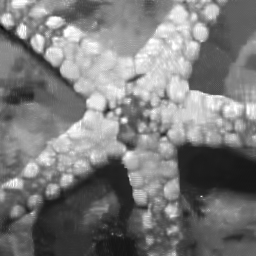}
        \end{minipage}
        \hfill
        \begin{minipage}[c]{0.18\textwidth}
        \includegraphics[width=\textwidth]{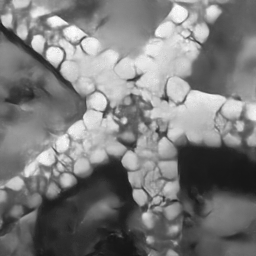}
        \end{minipage}
        \hfill
        \begin{minipage}[c]{0.18\textwidth}
        \includegraphics[width=\textwidth]{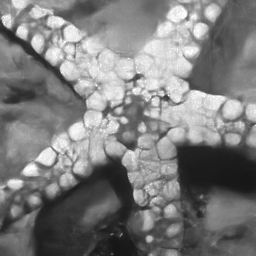}
        \end{minipage}
    \end{minipage}\\
    \vspace{-0.2cm}
    \caption{Synthetic images: Noisy, PPB (21.13 dB), SAR-BM3D (22.71 dB), SAR-CNN (23.37 dB), our method (23.32 dB).}

  \label{fig:zoom_images}
\end{figure*}

\begin{figure*}[t]
  \centering
    \begin{minipage}[b]{\textwidth}
        \begin{minipage}[c]{0.18\textwidth}
        \includegraphics[width=\textwidth]{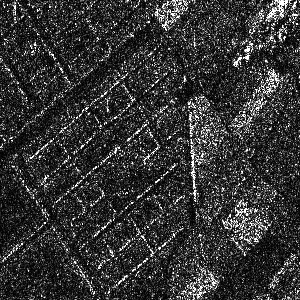}
        \end{minipage}
        \hfill
        \begin{minipage}[c]{0.18\textwidth}
        \includegraphics[width=\textwidth]{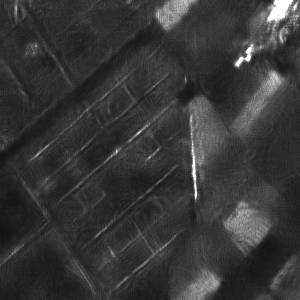}
        \end{minipage}
        \hfill
        \begin{minipage}[c]{0.18\textwidth}
        \includegraphics[width=\textwidth]{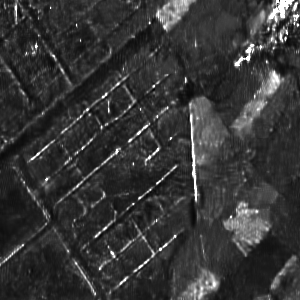}
        \end{minipage}
        \hfill
        \begin{minipage}[c]{0.18\textwidth}
        \includegraphics[width=\textwidth]{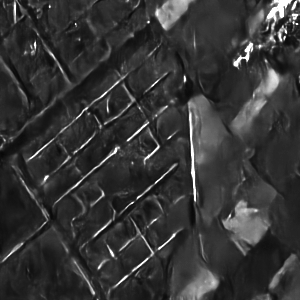}
        \end{minipage}
        \hfill
        \begin{minipage}[c]{0.18\textwidth}
        \includegraphics[width=\textwidth]{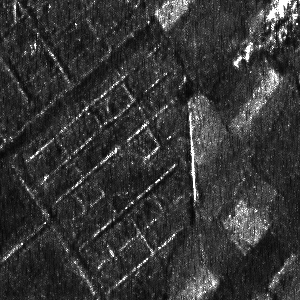}
        \end{minipage}
    \end{minipage}\\
    \vspace{-0.2cm}
    \caption{Real SAR images: Noisy, PPB, SAR-BM3D, SAR-CNN, our method.}
  \label{fig:real_images}
  \vspace{-0.6cm}
\end{figure*}

\vspace*{-8pt}
\section{EXPERIMENTAL RESULTS AND DISCUSSIONS}
\label{sec:experiment}
\vspace*{-8pt}

In this section we describe the results of our method through a two-step validation analysis. First, we train and test the network on a synthetic dataset where the availability of ground truth images allows to compute objective performance metrics. We compare our method with the following despeckling algorithms: PPB \cite{5196737}, SAR-BM3D \cite{5989862} and SAR-CNN \cite{CozzolinoCNN}. This allows to understand the denoising capability of our self-supervised method in comparison with both traditional methods and a CNN-based one with supervised training.
In the second experiment, training is conducted directly on real SAR images. To compare the despeckling methods, we rely on some no-reference performance metrics such as equivalent number of looks (ENL), and moments of the ratio image ($\mu_r$, $\sigma_r$), and on visual inspection.

The network architecture we use in the experiments is composed of four branches with shared parameters (handling the four directions of the blind-spot receptive field, see \cite{laine2019high}) in a first part with 17 blocks composed of 2D convolution with $3\times 3$ kernel, batch normalization and Leaky ReLU nonlinearity. After that, the branches are merged with a series of three $1\times 1$ convolutions.  

\vspace*{-8pt}
\subsection{Synthetic dataset}
\label{sec:synth}
\vspace*{-4pt}
In this experiment we employ natural images to construct a synthetic SAR-like dataset. Pairs of noisy and clean images are built by generating speckle to simulate a single-look intensity image ($L=1$).
During training patches are extracted from 450 different images of the Berkeley Segmentation Dataset (BSD) \cite{MartinFTM01}. The network has been trained for around 400 epochs with a batch size of 16 and learning rate equal to $10^{-5}$ with the Adam optimizer.
Table \ref{table:synth_images} shows performance results on a set of well-known testing images in terms of PSNR. It can be noticed that our self-supervised method outperforms PPB and SAR-BM3D. Moreover, it is interesting to notice that while the proposed approach does not use the clean data for training, it achieves comparable results with respect to the supervised SAR-CNN method.
Fig. \ref{fig:zoom_images} shows that also from a qualitative perspective. Despite the absence of the true clean images during training, our method produces images as visually pleasing as those produced by SAR-CNN with comparable edge-preservation capabilities.

\vspace*{-10pt}
\subsection{TerraSAR-X dataset}
\label{ssec:real}
\vspace*{-6pt}
In this experiment we employ single-look TerraSAR-X images\footnote{\url{https://tpm-ds.eo.esa.int/oads/access/collection/TerraSAR-X/tree}}.
Most of the despeckling works in literature assume the multiplicative speckle noise to be a white process. However, the transfer function of SAR acquisition systems can introduce a statistical correlation across pixels. One of the assumption for the blind-spot network training to work is that the noise has to be pixel-wise independent so that the network cannot predict the noise component from the receptive field.
Hence, both training and testing images are pre-processed through a blind speckle decorrelator \cite{6487399} to whiten them.
During training patches are extracted from 16000 $256 \times 256$ whitened SAR images.
The network has been trained for around 100 epochs with a batch size of 16 and learning rate of $10^{-5}$ with the Adam optimizer.

Table \ref{table:real_images} and Fig. \ref{fig:real_images} show the results obtained on three $1000 \times 1000$ test images disjoint from the training ones. ENL is computed over manually-selected homogeneous areas.
It can be noticed that the proposed method is very close to the desired statistics of the ratio image, showing that indeed it removes a significant noise component, and that it better preserves edges and fine textures. It also does not hallucinate artifacts over homogeneous regions, while SAR-CNN tends to oversmooth and produce cartoon-like edges. However, the degree of smoothing over homogeneous areas is somewhat limited as confirmed by the ENL values and deserves further investigation. We conjecture that residual spatial correlation in the speckle may affect the network on real images, since excellent performance is observed on synthetic speckle.

\vspace*{-10pt}
\section{Conclusion}
\label{sec:conclusion}
\vspace*{-6pt}
In this paper we introduced the first self-supervised deep learning SAR despeckling method which only requires real single look complex images. Learning directly from the true SAR data rather than simulated imagery avoids transfering between domains for improved fidelity. 

%


\end{document}